\documentclass[fleqn,10pt]{wlscirep}
\usepackage[utf8]{inputenc}
\usepackage[T1]{fontenc}
\usepackage{bm}
\title{GUIDe: Generative and Uncertainty-Informed Inverse Design for On-Demand Nonlinear Functional Responses}

\author[1,*]{Haoxuan Dylan Mu}
\author[1]{Mingjian Tang}
\author[1,2]{Wei Gao}
\author[1,*]{Wei ``Wayne'' Chen}
\affil[1]{J. Mike Walker \textquotesingle 66 Department of Mechanical Engineering, Texas A\&M University, College Station, United States}
\affil[2]{Department of Materials Science \& Engineering, Texas A\&M University, College Station, United States}

\affil[*]{hmu2718@tamu.edu, w.chen@tamu.edu}

\begin{abstract}
    Inverse design is a common yet challenging engineering problem, particularly for nonlinear functional responses such as mechanical behavior or spectral analysis. Deep generative models are motivated by intractability, non-existence or non-uniqueness of solutions, and the need for rapid solution-space exploration. In this study, we show that deep generative model-based and optimization-based approaches can provide incomplete solutions or hallucinate given out-of-distribution targets. To address this, we propose the Generative and Uncertainty-informed Inverse Design (GUIDe) framework, which leverages probabilistic machine learning, statistical inference, and Markov chain Monte Carlo to generate designs with targeted nonlinear behaviors. Instead of inverse mappings $\textbf{response} \mapsto \textbf{design}$, GUIDe adopts $\textbf{design} \mapsto \textbf{response}$: a forward model predicts each design's nonlinear functional response and evaluates the confidence under a user-specified tolerance. Sampling the solution space by this confidence yields diverse feasible designs. Our validation on nacre-inspired composites finds solutions beyond the training range, even under out-of-distribution targets.

\end{abstract}

\begin{document}
\flushbottom
\maketitle
\thispagestyle{empty}

\section{Introduction}\label{sec1}

\begin{figure}[htbp!]
    \centering
    \includegraphics[width=\linewidth]{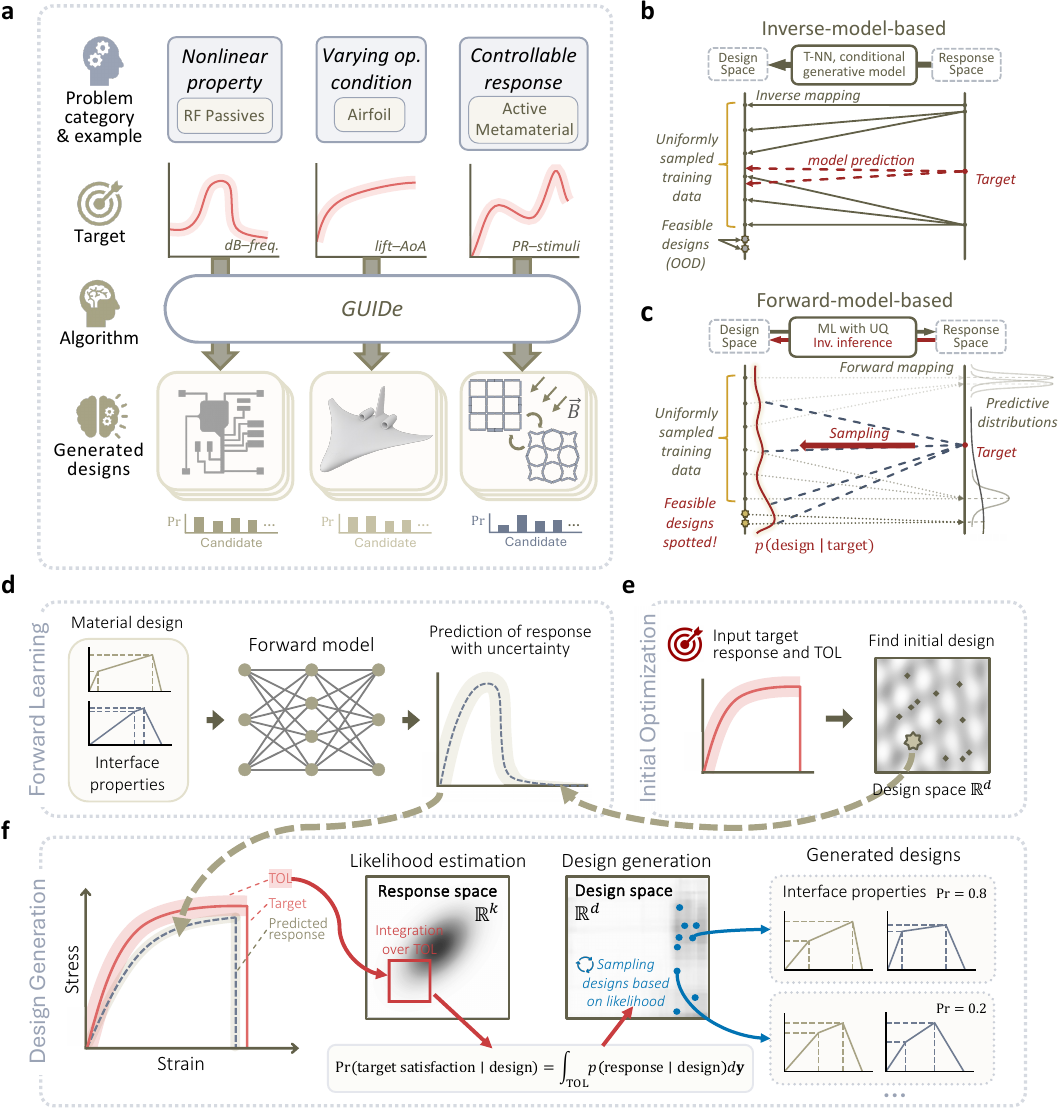}
    \caption{
        \textbf{a}, GUIDe applies to various inverse design problems where the targets are nonlinear functional responses. 
        \textbf{b}, \textbf{c}, a schematic comparison between inverse-model-based and forward-model-based (GUIDe) approaches. In both cases, the machine learning models are trained on uniformly sampled designs, with mappings between design and response spaces shown as arrows, depicting a scenario in which both approaches conditioned the same OOD target. \textbf{b}, Inverse-model-based algorithms (e.g., tandem neural network and conditional generative models) directly map responses to designs, but yield unreliable solutions for OOD targets.  \textbf{c}, Forward-model-based approaches predict responses with uncertainty, then the solutions are sampled based on their likelihood of meeting the target, highlighting the improved trustworthiness that distinguishes the proposed method. \textbf{d-f}, Architecture of GUIDe: \textbf{d}, A data-driven surrogate with uncertainty quantification is first trained to model the responses. \textbf{e}, Given a target response and the tolerance, stochastic optimization methods are applied to approach the feasible region by identifying an initial design. \textbf{f}, The design candidates are generated by sampling the posterior of the design parameters through MCMC, where the likelihood is approximated by integrating the predictive response space over the tolerance bound.
    }
    \label{fig:1}
\end{figure}

The problem of inverse design for functional responses aims to find a set of design solutions that achieves a target functional response, as depicted by Fig.~\ref{fig:1}a. It is a crucial topic across many engineering domains such as mechanics of materials\cite{ma2022deep, maurizi2025designing, zeng2023inverse, zhao2025extreme, wang2014design, filipov2015origami}, bioengineering\cite{ni2024forcegen, paige2011rna, huang2016design, wegst2015bioinspired, beedle2023role}, acoustics\cite{dong2024inverse, popa2014non, chen2024generative}, photonics\cite{molesky2018inverse, liu2021tackling, liu2018generative, hughes2018adjoint, raju2022maximized, tanriover2022deep}, electromagnetics\cite{karahan2024deep, sanjari2025reconfigurable, wang2021inverse}, and aerospace engineering\cite{sekar2019inverse, lei2021deep, yang2023inverse}. These applications often involve design for responses in various forms, such as nonlinear properties (e.g., material's stress--strain curve\cite{deng2022inverse} and gain--frequency relation of RF passives\cite{karahan2024deep}), response under varying operation conditions (e.g., airfoil’s lift coefficient under different angles of attack\cite{qu2015airfoil}), and controllable functional response (e.g., tunable Poisson’s ratio of active metamaterials under external stimuli\cite{ma2022deep}). While the advancing tools of optimization algorithms and scientific machine learning (SciML) techniques\cite{guo2021artificial, genty2021machine, so2020deep, lee2024data, brunton2021data} pushes the boundaries of broader applications, many fundamental challenges persist in inverse design, including the difficulty of learning an ill-conditioned inverse mapping, the non-existence or non-uniqueness of solutions, constrained data resources, uncertainty quantification, out-of-distribution (OOD) generalization, and 
modeling a highly non-convex and nonlinear design-response relationship.

For instance, a common approach for inverse design is gradient-based optimization, such as topology optimization\cite{xia2017recent, sigmund2013topology,white2019multiscale}. However, under nonlinear targets, the objective function often exhibits multiple local optima separated by steep energy barriers, making the exploration of the design space highly challenging. To address this, metaheuristics, such as evolutionary strategies\cite{beyer2002evolution, wierstra2014natural, mirjalili2019genetic, lambora2019genetic} and swarm intelligence\cite{wang2018particle} have been implemented, usually boosted by surrogate modeling approaches to reduce computational cost\cite{deng2022inverse}. Although metaheuristics are strong at global search, they can suffer from premature convergence without diversity-preserving mechanisms, collapsing into neighborhoods of local near-optima and thereby limiting the discovery of diverse design candidates. As a result, the quality of the outcomes is sensitive to the predictive accuracy of the surrogate model and often relies on expensive validation procedures.

Another generally applied approach for inverse design is Bayesian optimization (BO)\cite{ZHANG2025102359, park2023multi, xue2020machine}, which utilizes a probabilistic surrogate, typically a Gaussian process, to approximate the objective and quantify uncertainty, and then uses an acquisition function to balance exploration and exploitation. Its key limitation for on-demand inverse design is the need to acquire new data through expensive evaluations (e.g., simulations, experiments, or human annotations) and retrain the surrogate each iteration---this prevents fast design space exploration. While recent advances in in-context learning employ pre-trained foundation models as surrogates\cite{muller2023pfns4bo, yu2025git, rakotoarison2024context}---thereby eliminating iterative retraining and substantially reducing the computational budget---the overall workflow still depends on sequential data acquisition, which remains time-consuming and resource-intensive. Moreover, for different design tasks targeting distinct response behaviors, the process must be re-initialized with a newly sampled dataset tailored to each task, incurring high long-term cost.

More recently, data-driven approaches, particularly those leveraging deep learning, have emerged as compelling alternatives for inverse design: techniques such as tandem neural networks\cite{liu2018training, bastek2022inverting} and conditional generative models\cite{bastek2023inverse, ma2019probabilistic, trippe2022diffusion, wang2022ih, wang2024diffmat, yang2021deep, ma2022pushing, jiang2019free, seo2025physics}, which take target responses as input and generate corresponding designs via deep neural networks in a single shot, have shown strong potential in modeling intricate and highly-coupled relationships between design and response spaces. However, despite their promise, these methods often suffer from high data demands, a prolonged training process involving exhaustive hyperparameter tuning, and, most notably, facing challenges in modeling the ``\(\text{response} \mapsto \text{design}\)'' inverse mapping.

\begin{figure}[htbp]
    \centering
    \includegraphics[width=0.75\linewidth]{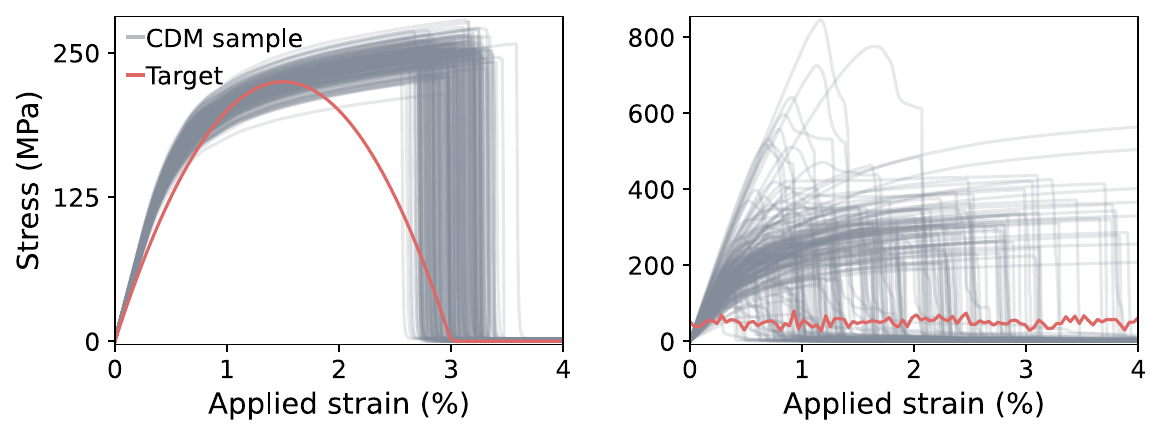}
    \caption{
        The design results obtained from the conditional diffusion model given OOD targets (i.e., stress–strain curves represented by a parabola and by Gaussian noise). The results reveal that CDM can generate irrelevant solutions (detailed in Sec.~\ref{subsubsec2}). On the other hand, GUIDe identified low likelihood of feasible solutions and avoided sampling misleading designs.
    }
    \label{fig:2}
\end{figure}

Especially for generative models, the challenges of inverse mapping arise from non-existence or non-uniqueness of solutions, ill-conditioning of inverse problems, lack of reliable uncertainty quantification, and difficulty of learning a robust mapping from sparse or ambiguous training data. For most of the inverse-mapping approaches, design confidence and uncertainty are hard to obtain because the conditional distribution of design is learned without a tractable likelihood, which tends to underrepresent multimodality and epistemic uncertainty, and predictions are frequently overconfident under OOD conditions. Fig.~\ref{fig:1}b presents a schematic of the inverse-model-based approach, which is trained on uniformly sampled design data. As uniform design sampling usually does not guarantee a full response space coverage, we consider a case where the target is underrepresented. In this setting, the inverse model will predict unreliable solutions. This behavior is demonstrated by a materials design case study shown in Fig. \ref{fig:2}---targeted on stress--strain curves given by a parabola and by Gaussian noise, a conditional denoising diffusion probabilistic model (DDPM)\cite{ho2020denoising} generates irrelevant solutions, as the target curves deviate too much from those in the training data (see Sec.~\ref{subsec4} for details).
This reveals that generating designs without diagnostic feedback (i.e., confidence or uncertainty measures) can significantly hamper the assessment of their trustworthiness. This issue is particularly critical in high-stakes applications that demand high confidence in generated solutions but face prohibitive costs for experimental validation.

To address the abovementioned issue, this work introduces the Generative and Uncertainty-informed Inverse Design (GUIDe) framework. Given a design target, GUIDe can generate a set of solutions based on their likelihood of meeting the target. In contrast to commonly used inverse-model-based approaches, GUIDe performs inference through a forward model, therefore bypassing the need to construct the more complex inverse mapping. Our prior work\cite{chen2024generative} explored a similar idea using random forest-based models and considering qualitative responses. Here, we propose a generalized framework that can be model-agnostic and applies to quantitative responses. As illustrated in Fig.~\ref{fig:1}c, GUIDe leverages the probabilistic predictive capacity of a data-driven forward model, which is trained to map design parameters to their associated response behaviors with uncertainty. Then with the forward prediction and a user-defined tolerance input, GUIDe evaluates the likelihood of a design meeting the target response, serving as a quantitative evaluation of design trustworthiness. To generate new design solutions, the framework constructs the Bayesian posterior by combining the likelihood with a prior over the design space and samples from it using Markov chain Monte Carlo (MCMC). For the same OOD targets shown in Fig.~\ref{fig:2}, GUIDe identified the low likelihood of solutions and did not sample any design, avoiding the generation of misleading solutions.

We evaluate the effectiveness of this framework by applying it to the design of interface properties in nacre-inspired composite materials. This type of material features a brick and mortar structure consisting of a ceramic phase and a matrix interface (Fig.~\ref{fig:4}a). Our design goal is to achieve prescribed target stress–strain curves under tensile loading by designing the interface behavior modeled using trilinear constitutive laws for the soft matrices. Notably, GUIDe enables the discovery of feasible solutions for OOD targets and can identify a diverse set of designs meeting the same target beyond training data. This capability offers valuable physical insights into the underlying fracture mechanisms in nacre-inspired composites. 

In summary, our contributions are four-fold.
First, we unveil the trustworthy issue of common inverse design approaches.
Second, we propose GUIDe, a general, forward-model-based inverse design framework that is agnostic to machine learning models and MCMC kernel options. Compared to inverse-mapping approaches, GUIDe offers improved reliability, a simpler training process, and lower data requirements.
Third, our quantitative analysis reveals a strong correlation between GUIDe's likelihood estimates and the actual feasibility rate of its generated designs. This finding indicates that, by prioritising high-likelihood samples, one can reasonably expect higher feasibility, therefore offering informative prior knowledge for decision-making without requiring expensive feasibility validation.
Finally, we demonstrate that GUIDe can extrapolate beyond the training data and consistently generate feasible designs, outperforming both the genetic algorithm and the conditional diffusion model in terms of feasibility, novelty, and diversity of generated solutions in the context of nacre-inspired composite material design. 

\section{Results}\label{sec2}
In abstract terms, a functional response, $\mathbf{y}(s) = f(\mathbf{x}, s)$, depends on design parameters $\mathbf{x}\in\mathbb{R}^d$, and an independent variable  \(s\) (e.g., frequency in spectral responses or strain in stress--strain relations). Solving the inverse problem involves building a system that models the conditional distribution \(p(\mathbf{x}\mid\mathbf{y}=\mathbf{y}^*)\), where $\mathbf{y}^*$ denotes the target. As shown by Fig.~\ref{fig:1}d-f, GUIDe is composed of three main stages.
First, we train a forward model to predict the nonlinear behaviors with uncertainties. Second, given a user-specified target response and tolerance, it performs stochastic optimization to search for an initial design within the posterior support. Finally, the method estimates the likelihood of designs meeting the target in light of the prediction, and generates new design solutions by sampling from the posterior.

\subsection{Forward modeling}\label{subsec1}
The forward modeling stage approximates the true function \(f\) with a surrogate model \(f_{\text{pred}}\), introducing uncertainty into the prediction \(\hat{\mathbf{y}}(s) = f_{\text{pred}}(\mathbf{x}, s)\), which is treated as a random vector with \(\mathbb{E}[\hat{\mathbf{y}}(s)] = \boldsymbol{\mu}_{{\hat{\mathbf{y}}}}\) and \(\mathrm{Cov}[\hat{\mathbf{y}}(s)] = \Sigma_{\hat{\mathbf{y}}}\). In our case study, the forward model takes trilinear interface parameters as input and outputs the predictive distribution of the corresponding stress--strain response (Fig.~\ref{fig:1}d). Extensive candidates exist for the option of the forward model. Note that while this framework generally requires a function modeling that returns the mean and covariance predictions, it does not consequently limit the choice of the model within this scope. There exist a few approaches, such as deep kernel learning (DKL)\cite{wilson2016deep} and spectral-normalized neural Gaussian process (SNGP)\cite{liu2023simple}, that can scalably learn the predictive multivariate distribution, being normal choices under our framework. However, in terms of more widely-used probabilistic approaches like Bayesian neural networks (BNN)\cite{neal2012bayesian}, Monte Carlo (MC) dropout\cite{gal2016dropout}, or deep ensembles\cite{lakshminarayanan2017simple}, they treat predictive uncertainty as conditionally independent across the response sequence, despite being more architecture-agnostic, and therefore are not directly applicable. To resolve this limitation, we devise a simple but effective kernel covariance modeling approach to graft these models into our framework: the correlation between different response dimensions can be accounted for through a radial-based similarity function, as it is a general and versatile approach for pointwise covariance function structure modeling, given by
\begin{equation}
\Sigma_{\hat{\mathbf{y}},{ij}}(\gamma) =
\sigma_{\hat{\mathbf{y}},{i}} \sigma_{\hat{\mathbf{y}},{j}}
\exp\left(-\gamma \| s_i - s_j \|^2 \right),
\label{eq:1}
\end{equation}
where we denote the standard deviation at $s_i$ as $\sigma_{\hat{\mathbf{y}},{i}}$, and parameter $\gamma > 0$ controls the covariance decay rate. Empirically, we found that in the nacre material design setting,   $\gamma \approx 0.35$ is favorable for smoothing the prediction samples and generating high-quality designs. Here, we provide some intuition for choosing a good $\gamma$ based on prior data. For training response labels $Y_\text{tr} \in \mathbb{R}^{h\times k}$, where $h$ and $k$ respectively denote the number of samples and the dimensionality of each response, the covariance matrix of training responses is indicated by $\Sigma^\text{tr} \in \mathbb{R}^{k\times k}$; then $\gamma$ can be estimated as 
\begin{equation}
\gamma^\ast = \arg\min_{\gamma} \left\| \Sigma_{\hat{\mathbf{y}}}(\gamma) - \Sigma^\text{tr} \right\|_F,
\label{eq:2}
\end{equation}
where $\|\cdot\|_F$ is the Frobenius norm. This couples the predicted distribution with the covariance structure existing within the observation. Note that modeling the response with a joint probability density across response dimensions is essential due to the intrinsic correlations and continuity inherent in physical processes. Preserving the same structure in the predicted samples helps ensure that likelihood estimation is not unduly influenced by noise when the underlying covariance is not explicitly modeled.

\begin{figure}[htbp]
    \centering
    \includegraphics[width=0.733\linewidth]{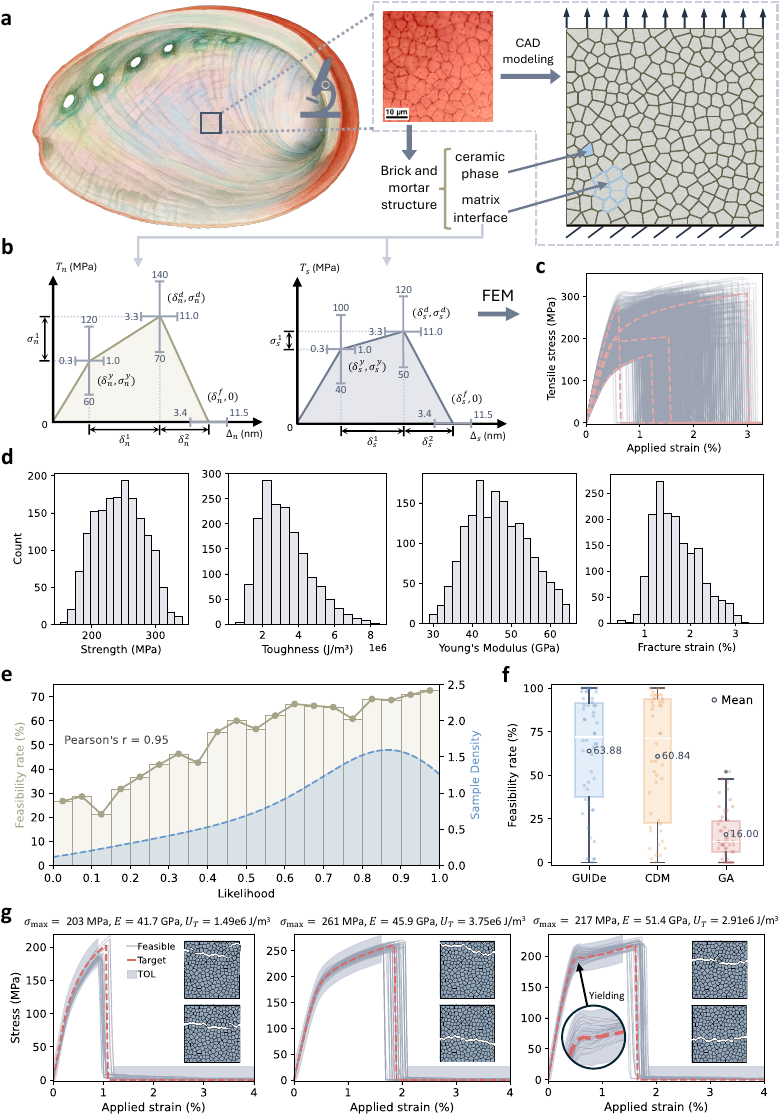}
    \caption{\textbf{a}, Schematic illustration showing the inside view of a red abalone shell. Inset shows a scanning electron microscopic top view revealing a Voronoi patterned brick-and-mortar structure (adapted with permission from ref.\cite{barthelat2007mechanics}, Elsevier Ltd.). A CAD model for a nacre-inspired composite is built to mimic this structure. \textbf{b}, Trilinear interface laws for normal and shear traction--separation relationship. Ranges of parameters in our dataset are marked by gray bars. \textbf{c}, Stress--strain curves from the dataset, computed via FEM under tensile loading. Representative response behaviors are labeled in red dashed lines. \textbf{d}, Distribution of strength, toughness, Young's modulus, and failure tensile strain for the stress--strain curves in \textbf{c}. \textbf{e}, Evaluation of the correlation between likelihood and actual feasibility rate. All samples were divided into 20 likelihood intervals, each containing at least 20 samples. The sample density is indicated by the blue dashed curve. A strong positive correlation can be observed between the likelihood score and the feasibility rate. \textbf{f}, Comparison of the feasibility rate distributions achieved by GUIDe, diffusion model, and genetic algorithm over 50 design scenarios. For each scenario, the target was sampled at random from the test dataset, and feasibility was computed from 50 generated designs. \textbf{g}, Inverse design results from GUIDe for three representative scenarios out of 50. For each scenario, two examples of the resulting fracture behavior are displayed.}
    \label{fig:4}
\end{figure}

\subsection{Likelihood estimation}\label{subsec2}

The likelihood of a design meeting the target depends on the response distribution \( p(\mathbf{y}(s) \mid \mathbf{x}) \). By defining the deviation from target as \(\boldsymbol{\delta}(s) = \mathbf{y}^*(s) - \mathbf{y}(s)\), a response is considered feasible when \( \|\boldsymbol{\delta}(s)\|_\infty \leq \boldsymbol{\epsilon} \), where $\|\boldsymbol{\delta}(s)\|_\infty = \max_i |\delta_i(s)|$, and $\boldsymbol{\epsilon}$ is a user-specified tolerance vector, optionally varying with $s$, that reflects strictness requirements. As indicated in Fig.~\ref{fig:1}f, the likelihood is formulated by the integration of the response distribution within tolerance: 
\begin{equation}
\mathcal{L}(\mathbf{x} \mid \mathbf{y}^*(s),\boldsymbol{\epsilon})\;=\;\Pr(\|\boldsymbol{\delta}(s)\|_\infty \leq \boldsymbol{\epsilon} \mid \mathbf{x}) \;=\;
\int_{a_1}^{b_1} \cdots \int_{a_k}^{b_k}
p(\boldsymbol{\delta}(s) \mid \mathbf{x}) \, d\mathbf{y},
\label{eq:3}
\end{equation}
where the integration bounds are defined as $\mathbf{a} = \mathbf{y}^*(s) - \boldsymbol{\mu}_{\hat{\mathbf{y}}} - \boldsymbol{\epsilon}$ and $\mathbf{b} = \mathbf{y}^*(s) - \boldsymbol{\mu}_{\hat{\mathbf{y}}} + \boldsymbol{\epsilon}$. When the response space is of low dimensionality (e.g., $k\leq10$), this likelihood probability can be computed using standard numerical quadrature methods such as Gaussian quadrature\cite{golub1969calculation} or adaptive multidimensional integration\cite{berntsen1991adaptive}. However, as the response exhibits increasing nonlinearity and is resolved over longer spatial or temporal domains, it is typically represented in a high-dimensional space to accurately capture fine-grained variations in behavior. Accordingly, the direct evaluation of integrals becomes computationally prohibitive due to the curse of dimensionality. To handle the high-dimensional multivariate normal integration in a controllable and efficient way, we applied a method of changing variables proposed by Genz\cite{genz2016numerical}, i.e., Cholesky factorization followed by an inverse-normal mapping. This cancels the Gaussian kernel with the Jacobian and recasts the problem as an integral on the unit hypercube, which we estimate efficiently with simple Monte Carlo. Implementation details are in Sec.~\ref{subsec9}. As a result, Eq.~\ref{eq:3} leads to a likelihood score in the range of $[0,1]$, reflecting the probability of a response meeting the target under ideal forward modeling assumptions. A synthetic example showing how sets of high-likelihood designs align with the real feasible domains is provided in Supplementary S1. 

\subsection{Design generation}\label{subsec3}
In high-dimensional settings, the predictive density of $\hat{\mathbf{y}}(s)$ is usually concentrated on a thin set. Therefore it is challenging to sample high-likelihood designs starting from random, especially under stricter error tolerances, as numerical precision limits cause likelihood values to underflow to zero in the heavy tail. To address this issue, we first utilize a stochastic optimization method to locate the support region (Fig.~\ref{fig:1}e). Specifically, the objective function is expressed as
\begin{equation}
\mathcal{S}(\mathbf{y}^*, \hat{\mathbf{y}}, \boldsymbol{\epsilon})
=
\underbrace{-\log\!\left( t + \frac{d_M(\mathbf{y}^*,\hat{\mathbf{y}})}{\sqrt{k}} \right)}_{\text{distance-based penalty}}
\;+\;
\underbrace{\frac{\lambda}{k}\sum_{u=1}^{k} \log\!\Big(\Phi(\xi_+^u)-\Phi(\xi_-^u)\Big)}_{\text{tolerance coverage approximation}}.
\label{eq:4}
\end{equation}
Here, with the stabilizer $t>0$, the first term is a monotone transform of the Mahalanobis distance between the predictive mean and the target; intuitively, it provides a smooth, covariance-aware score that pulls the predictive mean toward $\mathbf{y}^*$. The Mahalanobis distance is defined as
\begin{equation}
d_M(\mathbf{y}^*,\hat{\mathbf{y}})=
\sqrt{(\mathbf{y}^*-\boldsymbol{\mu}_{\hat{\mathbf{y}}})^{\!\top}\Sigma_{\hat{\mathbf{y}}}^{-1}(\mathbf{y}^*-\boldsymbol{\mu}_{\hat{\mathbf{y}}})}.
\label{eq:5}
\end{equation}
The second term is a z-score tolerance coverage estimation, in which we define
$\mathbf{r}=\mathrm{diag}(\Sigma_{\hat{\mathbf{y}}})^{-1/2}(\mathbf{y}^*-\boldsymbol{\mu}_{\hat{\mathbf{y}}})$,
$\boldsymbol{\tau}=\mathrm{diag}(\Sigma_{\hat{\mathbf{y}}})^{-1/2}\boldsymbol{\epsilon}$, and $\xi_\pm^u = r_u \pm \tau_u$. The per-dimension contribution is then evaluated independently as
$\Phi(\xi_+^u)-\Phi(\xi_-^u)$, i.e., the standard normal CDF over the interval $[\,\xi_-^u,\ \xi_+^u\,]$ in the standardized coordinates, with $\lambda$ as a tunable weight. Given that the z-score approximation ignores off-diagonal covariances, the second term alone can be spuriously increased by enlarging the marginal variances or by correlation patterns. Including the Mahalanobis term counters such effects by penalizing the discrepancy and supplies a stable signal in low-density areas where the approximate integral may flatten. Together, the two terms yield a stable surrogate when the true integral underflows, preserving a reliable ranking of designs and enabling fast convergence to the support. The initial sample is chosen as the first design showing a non-zero likelihood during the maximization of the objective. Once the starting sample is obtained, we employ MCMC to sample design candidates from the Bayesian posterior constructed through \(p(\mathbf{x} \mid \mathbf{y}^*(s),\boldsymbol{\epsilon}) \propto \mathcal{L}(\mathbf{x} \mid \mathbf{y}^*(s),\boldsymbol{\epsilon}) \cdot p(\mathbf{x})\), where the likelihood function \( \mathcal{L}(\mathbf{x} \mid \mathbf{y}^*(s),\boldsymbol{\epsilon}) = \Pr( \|\boldsymbol{\delta}(s)\|_\infty \leq \boldsymbol{\epsilon} \mid \mathbf{x})\) quantifies the probability of observing the target response given \( \mathbf{x} \). 

While optimization-based approaches guided by the objective function provide an efficient way to approach the regions of interest in the vast design space, it is important to notice that relying on optimization-based methods alone is insufficient to ensure comprehensive exploration of the design space. Our observations indicate that metaheuristic optimization algorithms such as genetic algorithm (GA)\cite{mitchell1998introduction, katoch2021review} and particle swarm optimization (PSO)\cite{kennedy1995particle, wang2018particle} often converge prematurely to a narrow region surrounding a single approximated global optimum. Although these global optimization strategies exhibit effectiveness in providing solutions that meet the target, this localized convergence limits their ability to identify multiple feasible regions, particularly in high-dimensional design spaces with multi-modal likelihood landscapes. Furthermore, reliance on a narrow solution distribution increases sensitivity to predictive inaccuracies of the forward model---small estimation errors at specific points may significantly compromise result quality, and often lead to adversarial minima, where excessive optimization exploits these inaccuracies to produce deceptively optimal yet poor-performing designs. This sampling-based method, coupled with likelihood estimation, demonstrates a stronger capacity for global exploration. Even in the presence of moderate approximation errors, it is capable of generating a diverse set of high-likelihood candidates. This robustness makes it particularly well-tailored for inverse design problems characterized by one-to-many mapping, where multiple distinct design configurations can yield similarly desirable target responses.

Another important requisite of a high-quality design generation involves selecting an appropriate MCMC kernel, which normally hinges on the posterior's dimension and geometry, the availability of derivative information, and the relative cost of each forward evaluation. With a moderately low-dimensional design space (e.g., $d\leq10$) and mild posterior correlations, a random-walk Metropolis-Hastings algorithm\cite{b1878965-730b-33c7-b1ad-dfd3acb6f61b}, preferably with an adaptive covariance such as the adaptive Metropolis scheme\cite{a5fbff14-79b7-3b59-a5f5-6424dbd2c77b}, or ensemble samplers\cite{goodman2010ensemble}, usually attain satisfactory effective sample sizes without the overhead of computing gradients. As dimensionality increases or when observing strong anisotropic correlations are present, random-walk proposals deteriorate with escalated integrated autocorrelation time, leading to slow exploration. In these settings, gradient-based MCMC---most notably Hamiltonian Monte Carlo (HMC)\cite{neal2011mcmc}, with the No-U-Turn Sampler (NUTS)\cite{hoffman2014no} as an adaptive trajectory-length variant---achieves higher sampling efficiency: simulating Hamiltonian dynamics yields long-distance proposals with high acceptance rate. These gains, however, assume the log likelihood is differentiable so that gradients are available---discrete parameters cannot be updated directly with HMC, and non-smooth or hard constraints can also break differentiability. In such cases, one typically marginalizes discrete quantities if possible or uses Metropolis-within-Gibbs updates for the discrete block interleaved with HMC for the continuous block. For potential design cases when the parameter dimension is itself variable, trans-dimensional methods such as reversible jump MCMC\cite{green1995reversible} should be considered. Moreover, severe multimodality can challenge all local kernels (transition strategies that rely on local proposals near the current state); here, tempering strategies, sequential Monte Carlo samplers\cite{del2006sequential}, or elliptical slice sampling\cite{murray2010elliptical} may help to approach geometric ergodicity. In brief, gradient-based samplers are often the method of choice for highly correlated, differentiable posteriors, while random-walk Metropolis remains efficient in low dimensions, especially when derivatives are unavailable, discrete variables dominate, or model evaluations are costly. 

\subsection{Inverse design of nacre-inspired material}\label{subsec4}

We applied our method to design nacre-inspired composite materials (see Supplementary S2 for background and motivation), modeling their stress--strain responses under longitudinal traction using the FEM. {As illustrated in Fig.~\ref{fig:4}a, the 2D FEM model is composed of isotropic hard grains, with zero-thickness cohesive interfaces separating the grains that govern the nonlinear mechanical response and failure of this material. We generated the geometry based on the Voronoi diagram, and adopted trilinear traction--separation laws to model the normal and shear responses of the interface. As shown in Fig.~\ref{fig:4}b, we introduce 10 independent design parameters to describe the interface laws: $\left\{\sigma_n^y, \Delta\sigma_n^1,\delta_n^y,\Delta\delta_n^1,\Delta\delta_n^2,\sigma_s^y, \Delta\sigma_s^1,\delta_s^y,\Delta\delta_s^1,\Delta\delta_s^2\right\}$, where the superscripts $y$ represent the yielding point, and the incremental parameters with $\Delta$ help track the state of damage evolution from the initiation to the complete failure. Details of the FEM configuration are in Sec.~\ref{subsec6}.
}

In this case study, we used long short-term memory (LSTM)\cite{hochreiter1997long} as the basic architecture to model the stress--strain evolution, with MC dropout to quantify the uncertainty. The training data consists of only 1,669 sets of design parameters and their corresponding stress--strain responses, while an additional set of 5,667 samples is split in half for validation and testing. 
The coverage of data (design parameters, responses, and derived properties) is illustrated in Fig.~\ref{fig:4}b--d and further discussed in Sec.~\ref{subsec5}. Particle swarm optimization (PSO) is applied to generate an initial sample for MCMC. As a population-based heuristic, PSO efficiently explores complex, non-convex landscapes and quickly approaches high-likelihood regions. Empirically, for this case study, PSO converges to a viable initial sample within a few iterations by maximizing Eq.~\ref{eq:4}. For PSO setup and parameterization details, see Sec.~\ref {subsec8}. Subsequently, we adopt random-walk Metropolis as the MCMC kernel, leveraging the fact that the proposal distribution is symmetric. 

The target satisfaction of generated solutions was validated using FEM. In the following discussion, we will refer to designs that achieve the target as “feasible designs”. {Notably, in damage simulation, FEM results depend on mesh discretization, step size, and solver tolerances; thus, even subtle setup differences for the same design can produce variability in the stress–strain response near the fracture onset and can alter the predicted damage evolution path. 
Therefore, we disregard the misalignment of the stress--strain curve within 13\% of the target fracture strain when determining feasibility. }

\begin{figure}
    \centering
    \includegraphics[width=0.7\linewidth]{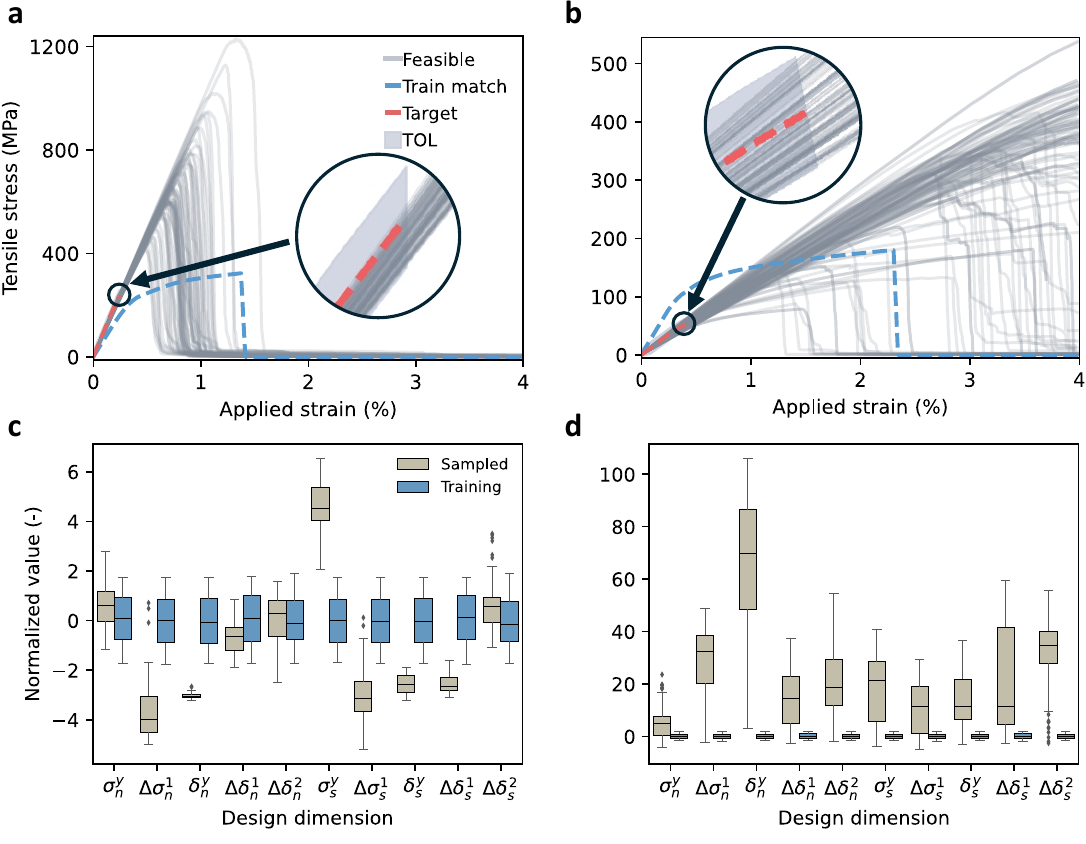}
    \caption{Design scenarios with OOD targets in the linear regime. \textbf{a}, \textbf{b}, Two design scenarios characterizing ultra-stiff (left) and ultra-soft (right) targets. Stress--strain curves (computed by FEM) for generated feasible designs are shown in comparison with the target, shaded tolerance range, and the closest match in the training data set. By adjusting the tolerance at different strain stages, the constraint was relaxed after the elastic regime; therefore, the model is free to explore designs with various post-yielding behavior. \textbf{c}, \textbf{d}, The comparison between the standardized range of training data and the feasible samples from GUIDe for design scenarios in \textbf{a} and \textbf{b}, respectively. It shows that GUIDe explores well beyond the bounds of the training data and still locates valid solutions.}
    \label{fig:5}
\end{figure}

\subsubsection{In-distribution targets}\label{subsubsec1}
To assess GUIDe’s generalization and its ability to rapidly generate feasible designs, we selected 50 responses from the test dataset as targets and specified a ±10\% tensile-strength tolerance for each. For every target, we discarded the initial 20 burn-in samples in MCMC, and retained the subsequent 50 designs---yielding 2,500 candidate designs in total. We benchmarked GUIDe against two inverse design baselines: (i) a surrogate model-assisted genetic algorithm (GA) that represents the prevalent forward-model-based strategies and (ii) a conditional diffusion model (CDM)\cite{zhang2023adding, 10.5555/3495724.3496298}, following the state-of-the-art architecture, as one of the leading inverse-model-based approaches as discussed in Sec.~\ref{sec1}. Configurations for the two baselines are provided in Supplementary S3 and S4. For every target, we draw 50 conditional samples. Fig.~\ref{fig:4}f suggests that GA is the least effective baseline: only 16\% of its sampled designs fall within the target tolerance. In comparison, GUIDe and the conditional diffusion model achieve average feasibility rates of 63.9\% and 60.8\%, respectively. A more nuanced picture emerges when likelihood scores are taken into account. Fig.~\ref{fig:4}e overlays sample density with likelihood and feasibility. It indicates a clear positive trend: designs assigned higher likelihood scores by GUIDe correspond to a higher probability of meeting the target, with a Pearson correlation coefficient (Pearson's r) of 0.95\footnote{The MCMC naturally proposes very few low-likelihood candidates. To ensure the coverage of all likelihood intervals, we supplemented the original 2,500 samples by drawing 10 additional test curves and forcing 20 designs to be generated in each likelihood interval.}. Kernel-density estimation shows that most probability mass lies in the high-likelihood region, in which feasibility can exceed 70\%, whereas low-likelihood regions are associated with markedly lower success rates. This strong likelihood–feasibility correlation provides a useful heuristic: by prioritizing high-likelihood samples, one can reasonably expect higher feasibility; therefore, the score offers informative prior knowledge for decision-making without requiring explicit feasibility validation.
Fig.~\ref{fig:4}g provides qualitative examples of GUIDe’s generated designs for three distinct targets (for more targets see Supplementary S5), each coupled with two example fracture patterns. The left panel depicts a brittle target that fails at around 1\% tensile strain, with GUIDe reproducing the abrupt drop in stress. The middle panel presents a tougher, stronger target, where the generated designs closely follow the higher peak stress and extended deformation. The right panel shows a curve with a clear yield point followed by a plateau-like toughening stage, and GUIDe effectively captures both the yield transition and the subsequent flat region.

\subsubsection{Out-of-distribution targets}\label{subsubsec2}
Beyond testing data, we devise four OOD design targets to evaluate GUIDe’s \textit{extrapolation} capability. For each design scenario, 2,000 design candidates are sampled. To balance the diversity and computational cost of material analysis, we select 200 designs from the sampled set, ensuring broad coverage of the design space using the max-min-distance criterion. Specifically, we choose the first point at random; thereafter, each new point is chosen to maximize its minimum Euclidean distance from the previously selected set. We then run FEM to test if each selected design meets the target and the pre-defined tolerance.
For the first target, we aim to explore designs with extremely high stiffness. Specifically, the target is 50\% stiffer than the stiffest training sample in the elastic region of $\varepsilon \in [0, 0.25]$, with a tolerance of 20 MPa. We set the tolerance to infinity after the elastic region. This gives GUIDe the freedom to explore diverse design solutions, provided they exhibit extremely high stiffness. In Fig.~\ref{fig:5}a, we present the described target in red and its closest match in the training data in blue. The tolerance is represented by the shaded region. The generated feasible stress--strain responses, depicted in gray curves, exhibit diverse mechanical behaviors. For example, their tensile strengths, ranging from 410.5 MPa to 1229.6 MPa, are 251\% higher than the maximum strength found in the training set. 
The second target is a response with initial tangent stiffness equal to half that of the softest training-set design. The tolerance in the elastic region (i.e., $\varepsilon \in [0, 0.4]$) is set to 10 MPa, while we remove tolerance constraints for $\varepsilon \in (0.4, 4]$. As shown in Fig.~\ref{fig:5}b, the model successfully identified feasible designs beyond the training set. These designs exhibit unexpected variability in mechanical behavior, with many featuring fracture points that extend beyond the strain range defined by the dataset. Fig.~\ref{fig:5}c and d present the comparison between the standardized range of training data and the feasible samples from GUIDe for the aforementioned two design scenarios. In both cases (especially the second case), GUIDe explores far beyond the bounds of the training data and still locates valid solutions.

\begin{figure}
    \centering
    \includegraphics[width=1\linewidth]{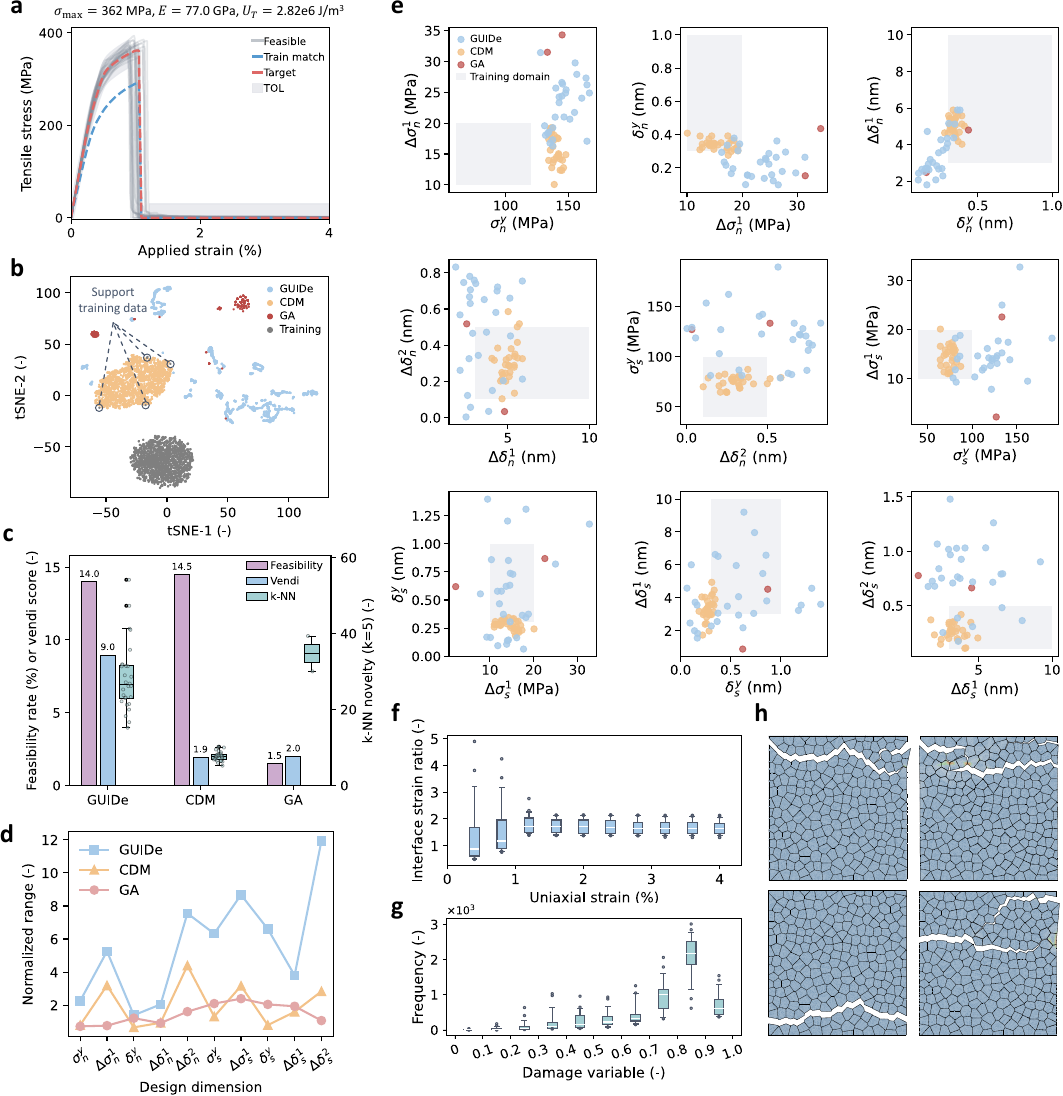}
    \caption{Design scenario with an OOD target featuring extremely strong and brittle mechanical behavior. \textbf{a}, Target response, closest matching solution in training data, and the feasible responses of designs generated by GUIDe. The target exhibits a tensile strength of 362 MPa, Young’s modulus of 77.0 GPa, and toughness of 2.82 MPa, while tolerance is as 8.3\% of $\sigma_\text{max}$. \textbf{b}, Design sample distribution of GUIDe, CDM, and GA, conditioned on the target introduced in \textbf{a}. GUIDe achieves the broadest exploration among the three. \textbf{c}, Quantitative evaluation of the feasible designs generated by the three methods in metrics of feasibility rate, Vendi score (diversity), and k-NN novelty. While CDM reaches a similar level of feasibility as GUIDe, the latter demonstrates superior performance in generating diverse and novel feasible designs. \textbf{d}, Standardized range of each design dimension across feasible designs obtained by the three methods. \textbf{e}, Feasible designs' distribution shown in pairwise scatter plots of each design dimension. GUIDe exhibits the widest coverage across the design space, often extending far beyond the training distribution. CDM samples are concentrated around the training prior, while GA demonstrates moderate exploration but identifies only three feasible designs. \textbf{f}, Distribution of interface-mode ratio $\phi$ versus applied strain from feasible designs generated by GUIDe. \textbf{g}, Distribution of frequencies across the damage variables of cohesive elements from feasible designs by GUIDe. \textbf{h}, Fracture patterns of four example feasible designs by GUIDe.}
    \label{fig:6}
\end{figure}

Our third extrapolation target is given by a nonlinear response characterizing a sharp post-fracture drop, shown in Fig.~\ref{fig:6}a. Starting from the origin, the stress initially increases in a near-linear manner, followed by nonlinear hardening after reaching the yield strength, and then an abrupt drop in stress, indicating fracture. We set the tolerance equal to 8.3\% (selected uniformly at random within 5–10\%) of the tensile strength of the target response. For this design scenario, the performance of GUIDe is again evaluated in comparison with GA and CDM. Fig.~\ref{fig:6}b shows an example of the t-SNE embedding of the 10-dimensional design parameter vectors for potentially feasible designs (within 10\% of the tensile strength of the target by prediction) among 2,000 solutions generated by GUIDe and GA, along with all the 2,000 CDM-generated designs and the training set; this figure describes how the design space is explored by three methods. GUIDe could discover several disconnected OOD regions away from the training manifold. Interestingly, although CDM samples occupy a cluster that is out of the manifold of the prior, it encloses a small subset of training points (highlighted with circles). These CDM-supporting training designs suggest that the diffusion model may primarily interpolate within the training data, rather than searching into novel design regions, which is consistent with the fact that the CDM is trained to approximate the seen distribution. GA designs usually overlap some of the GUIDe clusters but remain concentrated in limited sub-regions, indicating less coverage of the potentially feasible OOD space. As illustrated by Fig.~\ref{fig:6}c, we quantify the novelty of each feasible design among the three methods with k-NN distance\cite{cover1967nearest} ($k = 5$) to the training set. GUIDe achieves the highest novelty score, confirming its excellent extrapolation ability. This is followed by GA, whereas the conditional diffusion yields markedly lower novelty. The diversity was measured by the Vendi score\cite{friedman2022vendi}, a reference-free diversity metric for generative modeling through matrix-based entropy. GUIDe again outperforms the other two baselines by showing a 9.0 Vendi score versus 1.9 for CDM and GA, which could be explained by Fig.~\ref{fig:6}d as the feasible samples of GUIDe consistently span the broadest range in every design dimension. Fig.~\ref{fig:6}e shows the distribution of feasible designs pairwise in each design dimension (full distribution is presented in Supplementary S6). Here, the conditional diffusion model reaches a similar feasibility rate compared to GUIDe. In contrast, GA only generates three feasible designs out of 200. It can be interpreted that the GA is not robust to the prediction error of the forward model, especially given that the model was trained on fewer than 2,000 designs. Notably, these feasible solutions are not among the top-ranked candidates based on GA's fitness function (i.e., the MSE between the target and the predicted responses); instead, their predicted responses do not meet the target. This suggests that, in this case, the GA’s success in identifying feasible designs is largely due to luck. In addition, although the predicted responses of the top-ranked GA designs appear closely aligned with the target, none of their responses computed from FEM fall within the tolerance, indicating GA's vulnerability to prediction error. 

\begin{figure}[htbp!]
    \centering
    \includegraphics[width=1\linewidth]{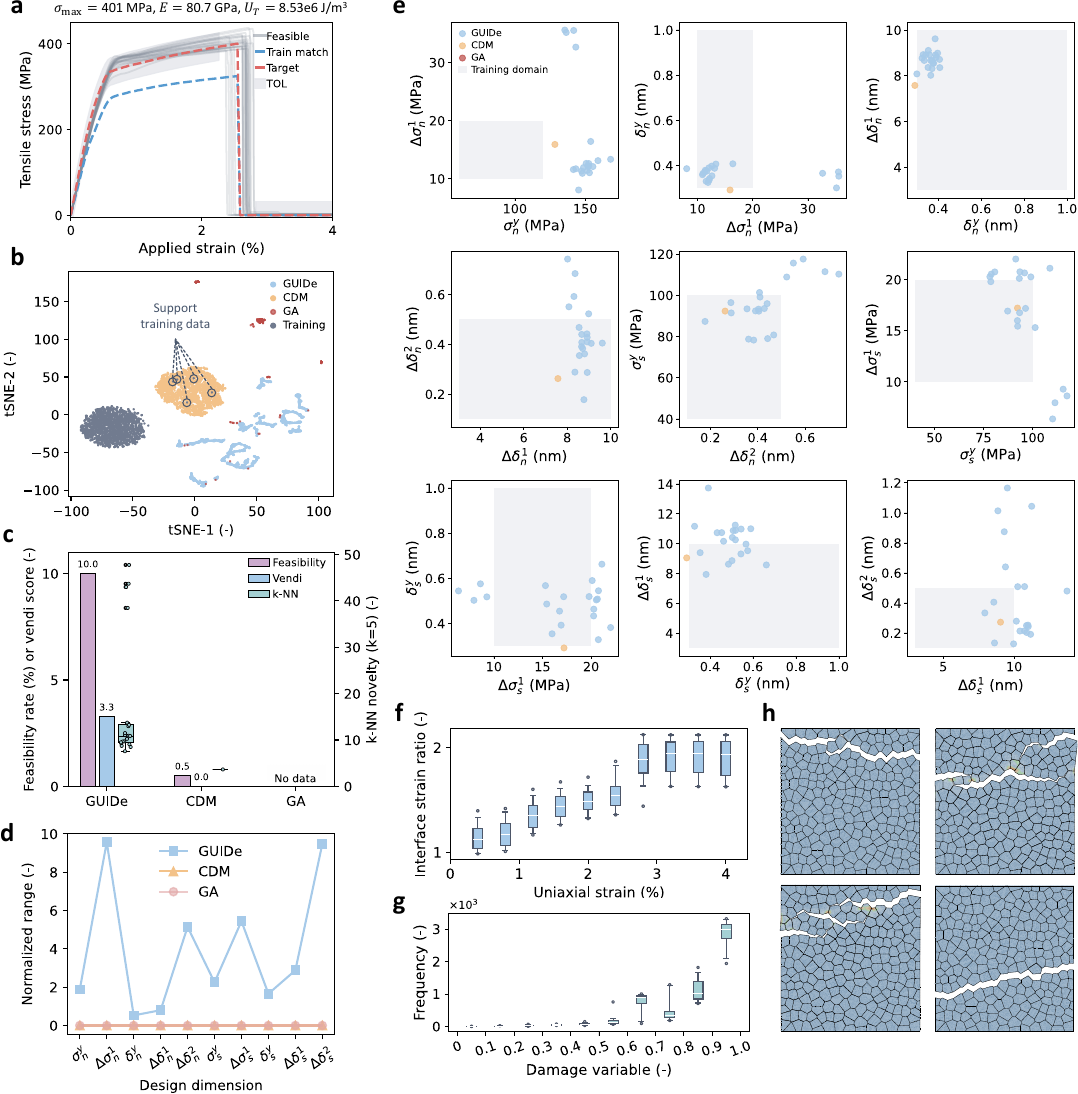}
    \caption{Design scenario with an OOD target featuring a plateau damage evolution. \textbf{a}, Target response, closest matching solution in training data, and the feasible responses of designs generated by GUIDe. The target exhibits a tensile strength of 401 MPa, Young’s modulus of 80.7 GPa, and toughness of 8.53 MPa, and tolerance is specified as 7.5\% of $\sigma_\text{max}$.
    \textbf{b}, Distribution of design samples generated by GUIDe, CDM, and GA, conditioned on the target in \textbf{a}. GUIDe achieves the broadest exploration among the three.
    \textbf{c}, Quantitative comparison of the feasible designs obtained from the three methods, evaluated by feasibility rate, Vendi score, and mean k-NN novelty. With the target being poorly represented in the training data, GUIDe demonstrates better performance over the benchmarks, excelling in the feasibility rate, diversity, and novelty of the discovered feasible designs.
    \textbf{d}, Standardized range of each design parameter among feasible samples generated by the three methods.
    \textbf{e}, Feasible designs' distribution shown in pairwise scatter plots across all design parameters. GUIDe exhibits the widest coverage across the design space, extending far beyond the training distribution. In contrast, CDM produces only one feasible design, while GA fails to identify any.
    \textbf{f}, Distribution of interface-mode ratio $\phi$ versus applied strain from feasible designs generated by GUIDe. 
    \textbf{g}, Distribution of frequencies across the damage variables of cohesive elements from feasible designs by GUIDe.
    \textbf{h}, Fracture patterns of four example feasible designs by GUIDe.}
    \label{fig:7}
\end{figure}

Fig.~\ref{fig:7} shows the fourth extrapolation case, where we examined a more exotic target with a plateau damage evolution followed by a sharp drop at failure. The tolerance is set to be 7.5\% of the target tensile strength. Results are illustrated using a similar set of figures (Fig.~\ref{fig:7}a--e). Different from the previous case, only one of the 200 CDM designs meets the target, leading to a k-NN novelty score of 3.7. Moreover, GA failed to discover designs that show prediction alignment, resulting in none of the generated designs meeting the target. In contrast, GUIDe successfully identifies feasible solutions outside the seen data distribution, achieving a 10\% feasibility rate, a 3.3 Vendi score, and a mean k-NN novelty of 17.0. According to Fig.~\ref{fig:7}b, the CDM-generated designs again remain clustered around some training samples, suggesting limited generalization to unexplored domains, while GUIDe offers a broad set of diverse candidates and successfully discovers multiple feasible regions.

To further study how the models behave under more extrapolative conditions, such as anomalous or physically unattainable targets, we formulate a parabola and a Gaussian noise as input response. In both scenarios, GUIDe refrains from generating designs, successfully preventing misleading outputs, owing to its OOD uncertainty recognition and probabilistic inference under tolerance constraints. Without uncertainty quantification, GA's top candidates have predictive results close to the target, but the real FEM response deviates. The diffusion model failed to detect such anomalies and continued to generate designs from the learned distribution. As shown in Fig.~\ref{fig:2}, the responses of all CDM-generated designs deviate from the specified targets---especially in the case of Gaussian noise, where the model produced designs whose responses exhibited a random, unstructured pattern. Further details of this design scenario can be found in Supplementary S7.

\subsubsection{Mechanics insights learned from inverse design}\label{subsubsec3}

There are two key observations emerging from the designs obtained by GUIDe, providing interesting mechanics insight into this interface-controlled deformation mechanism. 
First, for the third and fourth OOD design cases (hereafter referred to as ``Case 3'' and ``Case 4''), interface parameters associated with the yielding in the normal (opening) direction (i.e., $\sigma_n^y$, $\delta_n^y$, $\Delta\delta_n^1$) are much more tightly constrained by the target macroscopic stress–strain response than those associated with the shear (tangential) direction (i.e., $\sigma_s^y$, $\delta_s^y$, $\Delta\delta_s^1$). 
Although the specimen is loaded in uniaxial tension, this finding is not entirely intuitive: many interfaces are oblique to the loading direction and therefore experience mixed-mode (normal plus shear) separation. Nevertheless, the inverse design suggests the normal direction as having greater leverage on the global response.

To interpret this observation, we define an interface-mode ratio that aggregates the local kinematics over all interfaces,
\begin{equation}
\phi(\varepsilon)=\frac{\sum_{i}\varepsilon^{(i)}_{n}}{\sum_{i}\left|\varepsilon^{(i)}_{s}\right|},
\end{equation}
where $\varepsilon^{(i)}_{n}$ and $\varepsilon^{(i)}_{s}$ are the normal and shear strains on $i$-th interface at the applied macroscopic strain $\varepsilon$ (the absolute value prevents sign cancellation in shear). $\phi>1$ can be treated as normal-mode dominated and those with $\phi<1$ as shear-mode dominated. Box plots of $\phi$ versus applied strain across all learned designs (Figs.~\ref{fig:6}f and \ref{fig:7}f), including the interquartile range shown as blue boxes with the median marked as white lines, indicate that $\phi\approx 1$ at small strains with a wider spread (especially in Case~3), then increases during loading and converges to $\phi\simeq 1.7\text{--}1.8$ after failure. This result indicates that damage initiation and propagation are primarily governed by the normal opening of the interface law, while the shear component accommodates sliding but is less determinative of the macroscopic stress–strain response. This observation can be further confirmed by the fracture patterns in four different designs of two cases, where the main cracks propagate horizontally, as shown in Figs.~\ref{fig:6}h and \ref{fig:7}h.

The second observation is related to the interface cohesive energy, which is defined by the area under the interface traction–separation law, with normal and shear components (denoted $G_n$ and $G_s$).
Across the two design sets, Case 4 exhibits markedly narrower variation than Case 3:
$G_n\in[1.31,1.59]~\text{J/m}^2$ and $G_s\in[0.85,1.44]~\text{J/m}^2$ (Case 4) versus
$G_n\in[0.36,0.98]~\text{J/m}^2$ and $G_s\in[0.17,1.15]~\text{J/m}^2$ (Case 3).
The key difference in macroscopic behavior is that Case 4 shows an extended nonlinear regime in the stress–strain curve, where interfaces progressively damage before final rupture; by contrast, Case 3 ruptures at a smaller strain with limited nonlinearity. These results indicate that macroscopic responses featuring extensive damage evolution place tighter constraints on $G_n$ and $G_s$, even though the detailed shape of the traction–separation curve may vary.

To rationalize the second observation, we analyze the distribution of the cohesive damage variable $D$ ($D_i\in[0,1]$, 0: undamaged, 1: fully damaged) over all interface elements and all designs (Figs.~\ref{fig:6}g and \ref{fig:7}g). Because the stress--strain curve is prescribed, the macroscopic toughness or the total energy that must be dissipated is fixed; call this total $E_\text{tot}$. In our model, this dissipation is carried by the interfaces damage. Let $h_i$ be the per-element cohesive energy available at full failure, then an energy balance gives $E_\text{tot} \approx \sum_i D_i h_i$. 
In Case 4, the damage variable distribution is skewed toward larger values (predominantly D $>0.8$), so $\sum_i D_i \approx N$, where $N$ is the number of active interface elements. This reduces to $E_\text{tot} \approx N h$, if $h_i$ are similar for normal and shear, which means that once $E_\text{tot}$ is fixed by the stress--strain curve, the per-element dissipation $h$, and therefore $G_n$ and $G_s$, is effectively determined. This yields the narrow ranges observed in Case 4. 
By contrast, in Case 3 the damage distribution is more even with many mildly damaged elements ($D<0.5$), allowing trade-offs that satisfy the same $E_\text{tot}$ with either (i) many interfaces dissipating small amounts (small cohesive energy) or (ii) fewer interfaces dissipating larger amounts (large cohesive energy). This allows for wider $G_n$ and $G_s$ ranges in Case 3.

\section{Discussion}\label{sec3}
While advanced machine learning tools such as deep generative models have significantly accelerated and improved the accuracy of solving inverse design problems, core challenges remain. These include addressing data scarcity, ensuring the coverage of generated solutions, providing confidence and uncertainty estimations, and enhancing the generalization capabilities of the methods. GUIDe takes a step forward in tackling these challenges. The statistical inference and sampling in GUIDe allow for the exploration of the entire design space while considering model prediction uncertainty, yielding diverse design solutions coupled with confidence measures that better inform decision-making. In cases where the target response is OOD, GUIDe can still generate feasible designs based on its likelihood-awareness in potential solutions, whether within or beyond the training distribution. In extreme scenarios where the target response deviates too much from seen data, GUIDe refuses to generate any designs, preventing the output of unreliable or random solutions. Besides uncertainty quantification, the forward model-based paradigm further offers the potential for enhanced performance on OOD targets through the incorporation of physics-informed machine learning or causal models, an avenue we will explore in future research. Note that some methods for simulation-based inference (SBI)---which performs inference from simulator-generated data\cite{cranmer2020frontier}---solve inverse problems by training a surrogate model to learn the likelihood of a simulator's output data given input parameters (i.e., neural likelihood estimation), and then sampling the posterior. However, rather than requiring an explicit likelihood output, often costly and unstable, from the machine learning model, GUIDe computes a tolerance-aware feasibility probability between 0 and 1 based on the estimated full joint distribution of the high-dimensional response. The method incorporates design requirements, reduces data demand, and provides a more interpretable confidence measure for generated solutions with OOD detection.

In turn, given GUIDe's superior potential in addressing OOD targets and generating diverse solutions, we can leverage it to guide batch data acquisition in active learning and accelerate the discovery of novel designs with properties or performance far beyond those in the training data. Its likelihood estimation can be used to control the exploration-exploitation trade-off---higher likelihood favors feasible designs (exploitation), while lower likelihood sacrifices some feasibility for greater diversity (exploration).

Additionally, GUIDe’s modular architecture makes it compatible with a variety of machine-learning models and sampling strategies. This modularity makes it potentially widely applicable to problems with diverse design representations and varying complexities of design and response spaces. For example, by implementing gradient-informed MCMC kernels, such as Hamiltonian Monte Carlo approaches, the framework can generalize to more correlated and higher-dimensional design spaces.
Lastly, GUIDe allows for adjustable strictness of design targets by conditioning on the tolerance vector across different regions of the response. This provides the flexibility to weigh various parts of the response differently based on specific engineering requirements, without the need to retrain the model. This also enables the generation of designs with diverse behaviors that fulfill the same target. Together, these capabilities and potentials position GUIDe as a robust framework for generative inverse design.

\section{Materials and methods}\label{sec4}

\subsection{Dataset generation}\label{subsec5}
{To obtain a well-trained forward model for the inverse design, a broad spectrum of parameterized interface laws and their corresponding stress--strain relations were generated as the dataset. Fig. \ref{fig:4}b shows the range of 10 design parameters for the dataset. Among the parameters, each increment is constrained to be positive, and the hardening slope is capped by the elastic one, i.e., $\Delta\sigma_n^1/\Delta\delta_n^1<\sigma_n^y/\delta_n^y$ and $\Delta\sigma_s^1/\Delta\delta_s^1<\sigma_s^y/\delta_s^y$. Reducing stiffness reflects the evolution of micro-cracks and bridging mechanisms\cite{Li2005CST, Giusti2023Polymers}, and also guarantees positive energy dissipation together with numerical stability\cite{Park2011AMR} throughout our dataset construction and design generation processes. Using the trilinear laws leads to greater plastic deformation beyond the yielding point\cite{LINER2019349}. In contrast to the trapezoidal traction--separation law\cite{SALIH201695}, our formulation introduces an additional strengthening branch that enables further energy dissipation. These features contribute to a markedly nonlinear response of the material, making the inverse design more compelling and challenging.}

{In addition, each grain was treated as a linear-elastic solid with the Young’s modulus and Poisson's ratio of 10$^5$ MPa and 0.28, respectively. The high stiffness allows for neglecting the intragranular deformation and fracture. By randomly sampling the interface parameters among the ranges, we completed 1,669 FEM simulations under uniaxial loading, each producing a stress--strain curve. The curves (Fig. \ref{fig:4}c) cover the range of 150 to 350 MPa for the strength, and 1.0\% to 3.0\% for the failure tensile strain (Fig. \ref{fig:4}d), furnishing a sufficiently diverse training set for the forward model.}

\subsection{FEM setup}\label{subsec6}
{The FEM was performed through Abaqus/Standard 2023 (implicit solver) to simulate the uniaxial tension. Using the 2D nacre geometry in Fig.~\ref{fig:4}a, the model was meshed with quadrilateral-dominated plane-stress (CPS4R) elements for the hard grains and 4-node zero-thickness cohesive (COH2D4) elements for the interfaces. The interfaces followed the Maximum Nominal Stress (MAXS) damage initiation, with a tabular for normal and shear softening laws (including damage variables, displacements, and mode mix ratios) imported from Python for the damage evolution. Specifically, the extent of damage can be determined using a damage variable $D$ defined as follows, after reaching the yielding points in either direction:
\begin{equation}
    D=(1-\omega_1)D_n+\omega_1 D_s,\quad D_n=\frac{T_n\delta_n^y}{\Delta_n\sigma_n^y},\;D_s=\frac{T_s\delta_s^y}{\Delta_s\sigma_s^y},\quad \omega_1=\frac{T_s}{T_s+T_n},
\label{eq:7}
\end{equation}
where $D_n$, $D_s$ are the damage variables in the normal and shear directions, and $\omega_1$ is the mode mix ratio when determining the damage evolution in the FE software package. Notably, the normal and shear traction $T_n$, $T_s$ are used with their magnitude in the above expressions.
}

For the boundary conditions, the bottom edge was set to fixed, while the nodes on the top edge were kinematically coupled to a single reference point, applied with a uniformly distributed tensile displacement. The analysis used a single static step with automatic incrementation: the initial and maximum increments of 0.0001 and 0.01, and a total time period of 1.0.  We enhanced numerical stability by implicit viscous damping, whose factor was set to 10$^{-6}$. In our case, each simulation takes about 20--30 minutes on a commercial laptop CPU, and large studies were accelerated via parallel computing across multiple nodes on Linux high-performance computers (HPC).

\subsection{Forward model building and training}\label{subsec7}

The neural network was implemented using \texttt{TensorFlow}\cite{abadi2016tensorflow}~\texttt{v2.19.0}, comprising two stacked LSTM layers with 256 and 128 units, both employing a hyperbolic tangent activation, followed by two fully connected layers (256 units each) with linear activations. Dropout layers with rates of 0.2 are incorporated after each fully connected layer. The final output layer employs a softplus activation to ensure non-negative stress predictions. The input is represented as a tensor \( \mathbf{Q} \in \mathbb{R}^{100 \times 11} \), where each sample contains a sequence of $k=100$ steps at equally spaced strain levels from 0\% to 4\%. The $m$-th input at strain $\varepsilon_u$ is an 11-dimensional vector
\begin{equation}
\mathbf{q}_{m,u} = \big[ \varepsilon_u, \sigma_{n}^{y}, \Delta\sigma_{n}^{1}, \delta_{n}^{y}, \Delta\delta_{n}^{1}, \Delta\delta_{n}^{2}, \sigma_{s}^{y}, \Delta\sigma_{s}^{1}, \delta_{s}^{y},
\Delta\delta_{s}^{1}, \Delta\delta_{s}^{2}
\big]_m,
\label{eq:8}
\end{equation}
and each sequence maps to the corresponding stress vector, that is, \(\mathbf{y}_m=\big[y_1,\dots,y_k\big]_m^{\top}\). The model is trained using Adam\cite{kingma2014adam} with a batch size of 64, minimizing the loss function
\begin{equation}
\mathcal{L}_\text{f} = C\sum_{m=1}^{1669}\sum_{u=1}^{100}(\hat{y}_m(\mathbf{q}_{m,u})-y_{m,u})^2
\label{eq:9}
\end{equation}
where $y_{m,u}$ is the $u$-th component of $y_m$, and $C$ is the normalizing constant. Before training, we apply z-score normalization to standardize all the input features (i.e., design parameters and strains) following
\begin{equation}
\tilde{\mathbf{q}}_m = \frac{\mathbf{q}_m - \bar{\mathbf{q}}}{\boldsymbol{\sigma_\mathbf{q}}},
\label{eq:10}
\end{equation}
where the mean $\bar{\mathbf{q}}$ and the standard deviation $\boldsymbol{\sigma_\mathbf{q}}$ encompass the features within the whole data set. By re-enabling dropout during the prediction stage, probabilistic modeling is achieved through a set of stochastic forward passes, where each independent pass—characterized by different dropout masks—collectively forms an ensemble that approximates the predictive distribution of the stress--strain response for a given design parameter input. Formally, the predicted response follows a distribution
\(
\hat{\mathbf{y}}_\text{mcd} \sim \mathcal{N}(\boldsymbol{\mu}_{\hat{\mathbf{y}}}, \boldsymbol{\sigma}_{\hat{\mathbf{y}}}^2)
\), where the mean and the marginal variance are approximated as
\begin{equation}
\begin{aligned}
\boldsymbol{\mu}_{\hat{\mathbf{y}}} = \mathbb{E}[\hat{\mathbf{y}}] \approx \frac{1}{T} \sum_{t=1}^{T} \hat{\mathbf{y}}_\text{mcd}^{(t)}, \quad
\boldsymbol{\sigma}_{\hat{\mathbf{y}}}^2 \approx \frac{1}{T} \sum_{t=1}^{T} \bigl( \hat{\mathbf{y}}_\text{mcd}^{(t)} - \boldsymbol{\mu}_{\hat{\mathbf{y}}} \bigr) \odot \bigl( \hat{\mathbf{y}}_\text{mcd}^{(t)} - \boldsymbol{\mu}_{\hat{\mathbf{y}}} \bigr),
\end{aligned}
\label{eq:11}
\end{equation}
where \( \hat{\mathbf{y}}_\text{mcd}^{(t)} \) is the output of the \( t \)-th MC sample. To ensure adequate modeling accuracy for the distribution, we set $T=30$ so that the MC Dropout Gaussian model achieves a likelihood accuracy of at least $\beta = 0.95$, i.e.\ its geometric-mean likelihood on an independent test set is within $-\ln\beta = 0.051$ of the true generator’s value\cite{psutka2019sample, gal2016dropout}.

\subsection{PSO setup and parameter selection}\label{subsec8}

We employed a PSO algorithm implemented using a swarm of 80 particles following a star-topology communication scheme, in which each particle is attracted by the global best-performing particle. To guarantee a stable convergence, the PSO parameters are chosen according to the constriction factor approach by Eberhart\cite{eberhart2000comparing}. Specifically, the cognitive and social acceleration coefficients are both set to $c_1 = c_2 = 1.49445$, and the inertia weight is set to $w = 0.729$. The search space bounds were chosen as 
$\left( \max(0,\;\bar{\mathbf{x}}-\alpha\boldsymbol{\sigma}_{\mathbf{x}}),\;       \bar{\mathbf{x}}+\alpha\boldsymbol{\sigma}_{\mathbf{x}}\right)$, where $\bar{\mathbf{x}}$ is the sample mean of the design-parameter vector over the training set, and $\boldsymbol{\sigma}_{\mathbf{x}}$ is the variance. $\alpha$ denotes a scaling factor that determines the width of the search space, 
and we set $\alpha=4$ to enable the output initial sample to remain close to the training distribution 
while permitting moderate exploration. The truncation of the lower bound at zero ensures positive results, according to the fact that stress/strain variables cannot be negative.

\subsection{Likelihood evaluation over high-dimensional responses}\label{subsec9}

In the numerical computation of the probability under a $k$-variate normal distribution, the likelihood of design $\mathbf{x}$ given the target satisfaction is
\begin{equation}
\Pr(\|\boldsymbol{\delta}(s)\|_\infty \leq \boldsymbol{\epsilon} \mid \mathbf{x})
= \frac{1}{\sqrt{|\Sigma_{\hat{\mathbf{y}}}| (2\pi)^k}} 
\int_{a_1}^{b_1} \!\cdots \int_{a_k}^{b_k}
\exp\!\Bigl(-\tfrac{1}{2}\, \mathbf{y}^\mathsf{T} \Sigma_{\hat{\mathbf{y}}}^{-1} \mathbf{y} \Bigr)\, \mathrm{d}\mathbf{y}.
\label{eq:12}
\end{equation}
Let $\Sigma_{\hat{\mathbf{y}}} = L\,L^\top$ be the Cholesky factorization and set $\mathbf{y} = L\,\mathbf{y}'$.
Denoting the standard normal CDF by $\Phi(\cdot)$, define $y'_u = \Phi^{-1}(z_u)$ for $u=1,\dots,k$, with $z_u\in[0,1]$. Then
\begin{equation}
y_u = \sum_{v=1}^u l_{u,v}\,y'_v = \sum_{v=1}^u l_{u,v}\,\Phi^{-1}(z_v),
\label{eq:13}
\end{equation}
and the truncation $a_u \le y_u \le b_u$ induces the nested bounds
\[
d_1=\Phi\!\Bigl(\frac{a_1}{l_{11}}\Bigr),\quad e_1=\Phi\!\Bigl(\frac{b_1}{l_{11}}\Bigr),
\]
\begin{equation}
d_u(z_{1:u-1})=\Phi\!\Bigl(\frac{a_u - \sum_{v=1}^{u-1} l_{u,v}\,\Phi^{-1}(z_v)}{\,l_{u,u}\,}\Bigr),\qquad
e_u(z_{1:u-1})=\Phi\!\Bigl(\frac{b_u - \sum_{v=1}^{u-1} l_{u,v}\,\Phi^{-1}(z_v)}{\,l_{u,u}\,}\Bigr)\quad (u\ge2).
\label{eq:14}
\end{equation}
If $a_u=-\infty$ or $b_u=+\infty$, set $d_u=0$ or $e_u=1$, respectively. Because the Gaussian kernel and the Jacobian determinant cancel under the above change of variables, the integrand reduces to $1$ and
\begin{equation}
\Pr(\lvert \boldsymbol{\delta}(s) \rvert \leq \boldsymbol{\epsilon} \mid \mathbf{x})
=
\int_{z_1=d_1}^{e_1}
\int_{z_2=d_2(z_1)}^{e_2(z_1)}
\cdots
\int_{z_k=d_k(z_{1:k-1})}^{e_k(z_{1:k-1})}
1\,\mathrm{d}z_k\cdots \mathrm{d}z_1.
\label{eq:15}
\end{equation}
Equivalently, the final linear re-scaling
\begin{equation}
z_u \;=\; d_u(z_{1:u-1}) \;+\; w_u\,\bigl(e_u(z_{1:u-1})-d_u(z_{1:u-1})\bigr),\; w_u\in[0,1],
\label{eq:16}
\end{equation}
maps the domain to the unit hypercube $[0,1]^k$, with Jacobian $\prod_{u=1}^k\!\bigl(e_u-d_u\bigr)$, yielding
\begin{equation}
\Pr(\lvert \boldsymbol{\delta}(s) \rvert \leq \boldsymbol{\epsilon} \mid \mathbf{x})
=
\int_{[0,1]^k}
\prod_{u=1}^k \bigl(e_u(\mathbf{w}_{1:u-1})-d_u(\mathbf{w}_{1:u-1})\bigr)\,\mathrm{d}\mathbf{w}.
\label{eq:17}
\end{equation}
In this hypercube form, standard Monte Carlo is straightforward by sampling $\mathbf{w}\sim\mathrm{Unif}([0,1]^k)$. 

In certain design problems, the covariance matrix $\Sigma_{\hat{\mathbf{y}}}$ can become ill-conditioned---for example, when the response exhibits strong interdependencies, or when the predicted variance is extremely small or large at certain locations. Such situations may lead to numerical instability during Cholesky decomposition. To mitigate this issue, a diagonal jitter $\eta$ needs to be added to the covariance matrix, updating it as $\Sigma_{\hat{\mathbf{y}}} \leftarrow \Sigma_{\hat{\mathbf{y}}} + \eta I,\;\eta>0$. The choice of jitter should be large enough to meaningfully reduce the condition number of the covariance matrix, yet small enough to preserve its original structure and the encoded statistical information. A common criterion for selecting $\eta$ is based on the condition number threshold, empirically given by $\kappa(\Sigma_{\hat{\mathbf{y}}}, \eta) \cdot \varepsilon_{\text{mach}} \ll 1$, where the $\kappa(\cdot)$ denotes the condition number and $\varepsilon_{\text{mach}}$ is the machine epsilon.

\subsection{Design sampling strategy}\label{subsec10}

We employed the Metropolis-Hastings algorithm to sample feasible design configurations. At each iteration, a new proposal \(\mathbf{x}'\) is generated by adding Gaussian noise to the current design \(\mathbf{x}_c\):
\begin{equation}
\mathbf{x}' = \mathbf{x}_c + \boldsymbol{\zeta}, 
\quad \boldsymbol{\zeta} \sim \mathcal{N}(\mathbf{0}, \psi^2 \Sigma_q),
\label{eq:18}
\end{equation}
where the scaling factor \(\psi\) is set to be $2.38/\sqrt{d}$, which is asymptotically optimal for Gaussian random-walk Metropolis algorithms\cite{gelman1997weak}. The proposal covariance $\Sigma_q$ is quantified through the samples during the burn-in. To ensure the feasibility of proposed designs, the algorithm imposes the following constraints:
\begin{equation}
\mathbf{x}' \geq 0 \quad\text{and}\quad 
\frac{\Delta\sigma_n^1}{\Delta\delta_n^1} > \frac{\sigma_n^y}{\delta_n^y}, \quad
\frac{\Delta\sigma_s^1}{\Delta\delta_s^1} > \frac{\sigma_s^y}{\delta_s^y}.
\label{eq:19}
\end{equation}
If a proposal does not satisfy these physical consistency constraints, it is resampled until a valid design is obtained. The acceptance criterion follows the standard Metropolis-Hastings acceptance probability:
\begin{equation}
a(\mathbf{x}, \mathbf{x}') = \min\left(1,\;\frac{p(\mathbf{x}')\mathcal{L}(\mathbf{x}' \mid \mathbf{y}^*,\boldsymbol{\epsilon})/q(\mathbf{x}'\mid\mathbf{x})}{p(\mathbf{x})\mathcal{L}(\mathbf{x} \mid \mathbf{y}^*,\boldsymbol{\epsilon})/q(\mathbf{x}\mid\mathbf{x}')} \right),
\label{eq:20}
\end{equation}
where \(p(\mathbf{y}^* \mid \mathbf{x})\) denotes the predictive likelihood of the design \(\mathbf{x}\). With the assumption of a uniform prior and symmetric proposal, the function simplifies to:
\begin{equation}
a(\mathbf{x}, \mathbf{x}') = \min\left(1,\; \frac{\mathcal{L}(\mathbf{x}' \mid \mathbf{y}^*,\boldsymbol{\epsilon})}{\mathcal{L}(\mathbf{x} \mid \mathbf{y}^*,\boldsymbol{\epsilon})} \right).
\label{eq:21}
\end{equation}
Under this rule, proposals with a higher likelihood are always accepted; otherwise, they are accepted with probability equal to the likelihood ratio.

\section{Supplementary Information}\label{sec5}
The Supplementary Information is available at \url{https://doi.org/10.5281/zenodo.17291305}

\section{Acknowledgements}\label{sec6}
This work was supported by the startup funds from the J. Mike Walker ’66 Department of Mechanical Engineering at Texas A\&M University and the National Science Foundation (NSF) EDSE program (CMMI 2434393). We thank Dr. Wei Zhang for the support on the finite element analysis. Portions of this research were conducted with the advanced computing resources provided by Texas A\&M High Performance Research Computing. 

\bibliography{Bibliography}
\end{document}